# A Survey on Cross-Layer Design Frameworks for Multimedia Applications over Wireless Networks

Jaydip Sen and Shomik Bhattacharya
Convergence Innovation Lab, Tata Consultancy Services,
M2 & N2, Bengal Intelligence Park, Salt Lake Electronic Complex, Sector V, Kolkata, INDIA
Emails: {jaydip.sen, shomik.bhattacharya}@tcs.com

## Abstract

In the last few years, the Internet throughput, usage and reliability have increased almost exponentially. The introduction of broadband wireless mobile ad hoc networks (MANETs) and cellular networks together with increased computational power have opened the door for a new breed of applications to be created, namely real-time multimedia applications. Delivering real-time multimedia traffic over a complex network like the Internet is a particularly challenging task since these applications have strict quality –of-service (QoS) requirements on bandwidth, delay, and delay jitter. Traditional IP-based best effort service will not be able to meet these stringent requirements. The time-varying nature of wireless channels and resource constrained wireless devices make the problem even more difficult. To improve perceived media quality by end users over wireless Internet, QoS supports can be addressed in different layers, including application layer, transport layer and link layer.  Cross layer design is a well-known approach to achieve this adaptation. In cross-layer design, the challenges from the physical wireless medium and the QoS-demands from the applications are taken into account so that the rate, power, and coding at the physical layer can adapted to meet the requirements of the applications given the current channel and network conditions. A number of propositions for cross-layer designs exist in the literature.  In this paper, an extensive review has been made on these cross-layer architectures that combine the application-layer, transport layer and the link layer controls. Particularly the issues like channel estimation techniques, adaptive controls at the application and link layers for energy efficiency, priority based scheduling, transmission rate control at the transport layer, and adaptive automatic repeat request (ARQ) are discussed in detail.

## 1   Introduction

As the wireless networks evolved from circuit-switched voice traffic based 2G networks to an all-IP based packet-switched 3G network catering to a mix of high speed real-time traffic such as voice, multimedia teleconferencing, online gaming etc. and data-traffic such as WWW browsing, messaging, file transfers etc., there has been a dramatic change in the quality of service (QoS) requirements in terms of transmission accuracy, delay, jitter, throughput and so on. The 3G networks have already been deployed in many parts of Europe, Asia and America, and advanced research is going on for developing fourth-generation (4G) networks with a maximum bit rate of 100 Mbps or above. In order to achieve a successful and profitable commercial market for future wireless technology, network service designers and providers need to pay much attention to efficient utilization of radio resources due to fast growth of the wireless subscriber population, increasing demand for new mobile multimedia services and consequent diverse and more stringent QoS requirements. Traffic on wireless networks is becoming increasingly complex with a mix of real-time traffic such as voice, multimedia teleconferencing, and games, and data-traffic such as WWW browsing, messaging and file transfers etc. All these applications require widely varying and very diverse QoS guarantees for different types of traffic. Of late, various mechanisms have been proposed in the literature to support these QoS requirements. However, providing a robust QoS support for multimedia applications over wireless networks is a very challenging task due the following reasons [3].

- Different applications have different QoS requirements. Real-time media such as video and audio is delay-sensitive but capable of tolerating a certain degree of errors. Non-real time media such as web data is less delay-sensitive but requires reliable transmission.
- Wireless channels have high packet loss rate and bit error rate (BER) due to fading and multipath effects. Resulting packet loss and bit errors can have an adverse effect on the multimedia applications.
- Wireless channels have bandwidth limitation and fluctuations of the available bandwidth, packet loss rate, delay and jitter.
- Traditional transport layer protocols perform poorly in wireless networks since they assume congestion to be the primary cause for packet losses and unusual delay in the network. These protocols reduce the transmission rate whenever they observer packet loss. In wireless networks, the packet may be dropped due to channel errors, thereby resulting in unnecessary reduction in end-to-end throughput.
- The mobile devices are power constrained. Maintaining good media quality and minimizing average power consumption (for processing and communication) are two conflicting requirements.
- Receivers in multimedia delivery systems are quite different in terms of latency requirements, visual quality requirements, processing capabilities, power limitations, and bandwidth constraints. Moreover, multimedia may traverse different types of networks, e.g., wire-line networks, cellular networks, and wireless local area networks (WLAN). Each of these networks has different characteristics such as reliability, delay, jitter, bandwidth, and medium access control (MAC) mechanisms.

In view of the above constraints, a strict modularity and protocol layer independence of the traditional TCP/IP or OSI stack will lead to a sub-optimal performance of applications over IP-based wireless networks. For optimization, we require protocol architectures that require modification of the reference layered stack by allowing direct communication between protocols at non-adjacent layers or sharing state variables across different layers to achieve better performance. The objective of a cross layer design is to actively exploit this possible dependence between protocol layers to achieve performance gains. Although the cross layer design is an evolving area of research, considerable amount of work has already been done on this area. In this paper, we have made a survey of these works identifying their main contributions and areas in which those mechanisms are applicable.

The rest of the paper is organized as follows. Section 2 describes various QoS parameters that are relevant for multimedia applications. Section 3 depicts some generic cross-layer frameworks and identifies major areas and protocol layers in which cross-layer design can be applied for QoS support to multimedia applications. Section 4 describes the adaptations required in the link layer, the transport layer and the application layer of the standard protocol stack for an effective cross-layer design. Section 5 concludes the paper.

## 2 Different QoS Classes of Multimedia Applications

One major challenge in multimedia services over wireless networks is QoS provisioning with efficient resource utilization. Heterogeneous multimedia applications in future IP-based wireless networks require a more complex QoS model and more sophisticated management of scarce radio resources. QoS can be classified according to its implementation in the networks, based on a hierarchy of four different levels: bit, packet, call, and application [1]. Transmission accuracy, transmission rate (i.e., throughput), timeliness (i.e., delay and jitter), fairness, and user perceived quality are the main considerations in this classification:
• *Bit-level QoS* - To ensure some degree of transmission accuracy, a maximum BER for each user is required. Any transmission with BER greater than the maximum permissible limit is not acceptable for applications which have a stringent QoS requirement. Data applications are more sensitive to bit errors than video applications.

• *Packet-level QoS* – For delay-sensitive applications like voice over IP (VoIP) and videoconferencing, each packet should be transmitted within a delay bound. On the other hand, data applications like Internet downloads can tolerate delay to a certain degree. Throughput is a better QoS criterion for data applications. Each traffic type can also have a packet loss rate (PLR) requirement.

• *Call-level QoS* – Due to insufficient capacity at any given instant in a wireless system, there are chances that a new call will be blocked or a handoff call will be dropped. From the user's point of view, handoff call dropping is more unacceptable than new call blocking because the user might be in the middle of some important transaction during handoff. It is necessary to devise an effective call admission control to ensure new call blocking rather than dropping a handoff call in this kind of scenario.

• *Application-level QoS* - Application layer perceived QoS parameters like - the peak signal to noise ratio (PSNR) for video application and the end-to-end throughput for data application provided by the responsive TCP, more suitably represent the service quality seen by the end user, than bit and packet level QoS.

Another big challenge is to develop an accurate mapping mechanism for application layer QoS parameters to the lower layer (PHY layer) parameters so that the requirements specified at the application layer are suitably converted to the corresponding requirements in the lower layers before being passed over the carrier. Kumwilaisak et al have proposed one such mapping architecture [2]. In addition to QoS parameter mapping, an effective link layer packet scheduling scheme with appropriate power allocation is required to support bit- and packet- level QoS requirements of the applications running on mobile devices. Specifically, the power levels of the mobile devices should be managed in such a way that each mobile station (device) achieves the required bit energy to interference-plus-noise density ratio, and the transmission from /to all the mobile stations should be controlled to meet the delay, jitter, throughput, and PLR requirements.

## 3   Cross-Layer Design and Architecture

In this section, some of the cross layer design frameworks and algorithms currently existing in the literature are described in brief.  Salient features of some of these schemes are presented highlighting their contributions and areas of applications.

### 3.1   A Cross-Layer QoS Support Architecture for Multimedia over Wireless Networks

Zhang, Yang and Zhu have described a general architecture based on Universal Mobile Telecommunications System (UMTS) for multimedia delivery over wireless Internet [3]. Figure 1 depicts the architecture, where the multimedia server, base station (BS) (gateway) with media proxy, and heterogeneous mobile clients are deployed. Application-layer, transport-layer, and link-layer control mechanisms are all taken into account and suitably placed into this generic architecture, to achieve a good end-to-end quality of multimedia services.

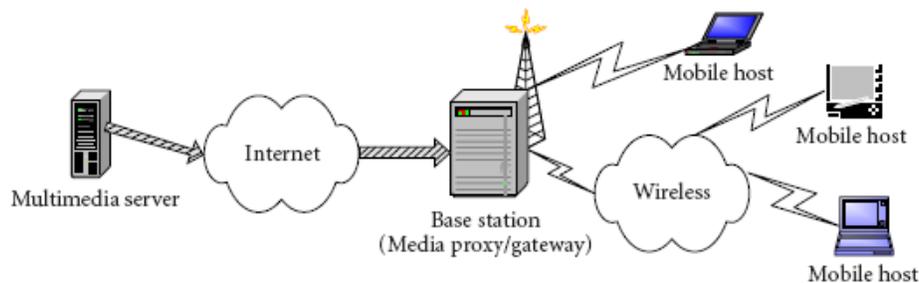

Figure 1: A general architecture for multimedia delivery over wireless Internet [3]

In Figure 1, the application is transmitted via TCP or UDP in the Internet part based on the traffic characteristics. IP packets arriving in the downlink to the UMTS network are transported to the radio network controller (RNC). Appropriate header compression techniques are applied to the packets in the packet data convergence protocol (PDCP) layer of the UMTS stack. The compression technique in the PDCP layer can vary depending on the implementation. The PDCP layer compresses the packets and attaches a header and further forwards them. It uses the service provided by a lower layer called radio link control (RLC) layer. The RLC layer is employed to support reliable upper layer protocols such as transmission control protocol (TCP). RLC uses sophisticated retransmission schemes to perform partial error recovery at the link layer, thus hiding transmission errors from upper layers and reducing the chances of performance degradation of upper layer protocols. The RLC protocol data units (PDU) of a particular IP connection are served by the MAC layer. In deterministic transmission time intervals (TTIs), the MAC layer entities ask the corresponding RLC layer entities for a certain number of RLC PDUs, which are then transferred through the radio interface in MAC frames. TTI refers to the length of an independently decodable transmission on the radio link. It is related to the size of the data blocks passed from the higher network layers to the radio link layer. In order to be able to adapt quickly to changing conditions in the radio link, shorter TTIs are preferable. However, in order to benefit more from the effect of interleaving and to increase the efficiency of error-correction and compression techniques, a system must have longer TTIs. Thus determination of TTI value is an optimization problem.

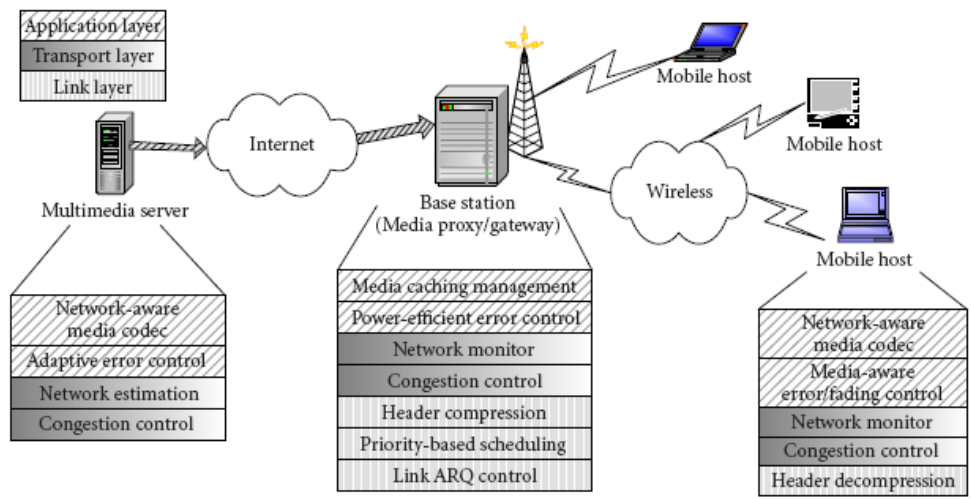

Figure 2. A cross-layer architecture for multimedia delivery over wireless Internet [3]

Figure 2 depicts the cross-layer architecture corresponding to the framework described in Figure 1. The authors have identified the following functionalities of the cross-layer architecture for QoS support to multimedia applications.

- *Dynamic wireless Internet condition estimation*: to track the varying wireless Internet conditions, network estimation mechanisms in different layers on the server, BS, and mobile-host side have to work together.
- *Network condition adaptation*: the cross-layer architecture has to adaptively adjust the amount of wireless Internet resources (i.e., the bandwidth) according to the varying network condition. This function is carried out by the congestion control module in the multimedia server and BS.
- *Network-aware media adaptation*: in response to the changing network conditions, media encoding mechanisms and different parts of media can be adaptively adjusted or tailored to maximize the system efficiency and perceived end-to-end delay.
- *Power efficiency and error robustness*: application and link-layer error control schemes can be used together for error robustness. Meanwhile, the overall power consumption in the mobile station (MS) should be minimized.

- *Efficient network utilization*: to improve the network utilization, especially in wireless channels, header compression is performed in both BS and mobile hosts.
- *Multi-services support*: for supporting multiple types of traffic having different types of QoS requirements, priority-based scheduling is an efficient way.
- *Network and client heterogeneity*: heterogeneity of different types of networks and client devices can be supported by QoS-adaptive proxy caching.

## 3.2  A Cross-Layer Resource Allocation Mechanism for 3G Wireless Networks

Jiang, Zhuang and Shen have proposed a cross layer design framework over a CDMA-based wireless network as depicted in Figure 3 [1]. In the proposed framework, three possible cross layer information flow have been identified., e.g., from physical to link layer, from link to transport layer and vice versa, and from link to application layer and vice versa. This leads to three cross-layer design approaches: (i) channel-aware scheduling, (ii) TCP over CDMA wireless links, and (iii) joint video source/channel coding and power allocation.

In channel-aware scheduling, the time-varying characteristics of a wireless channel are exploited by using a multiuser diversity framework to improve system performance. The principle of multi-user diversity is that for a cellular system with multiple mobile stations (MSs) having independent time-varying channels, it is very likely that there exists an MS with instantaneous received signal power close to its peak value. Overall resource utilization can be maximized by providing service at any time only to the MS with the highest instantaneous channel quality. The authors argue that with the capability to support simultaneous transmissions in a CDMA system, multi-user diversity can be employed more effectively and flexibly than traditional channel-aware scheduling schemes for a TDMA system. An MS does not need to wait until it has the best channel quality among all MSs, but rather it can transmit as long as its channel is good enough.

As a second cross layer design issue, an adaptive TCP has been proposed. Traditional TCP in wired networks adjusts its sending rate based on the estimated network congestion status so as to achieve congestion control or avoidance. In a wireless environment, TCP performance can be degraded severely as it interprets losses due to unreliable wireless transmissions as signs of network congestion and invokes unnecessary congestion control.  To improve TCP performance over wireless links, several solutions have been proposed to alleviate the effects of non-congestion-related packet losses [4]. The authors have argued that when a TCP connection is transmitted over CDMA cellular networks, apart from issues like congestion control, link errors etc, further considerations are to be made. First, CDMA capacity is interference-limited. TCP transmission from an MS generates interference to other MSs. It is desired to achieve acceptable TCP performance (e.g., a target throughput) and at the same time introduce minimum interference to other MSs (i.e. to require minimum lower-layer resources). Second, power allocation and control in CDMA can lead to a controllable BER, which affects TCP performance. Keeping in mind these issues, the authors have proposed an adaptive TCP that dynamically adjusts the sending rate of TCP segments (which will be fed back into the link layer transmission queue) according to network congestion status (e.g., packet loss and round-trip delay). A link layer design parameter ultimately determines the packet loss rate and transmission delay over the wireless link and therefore affects the TCP performance. With a proper choice of this link layer design parameter it will be possible to achieve the target TCP throughput.

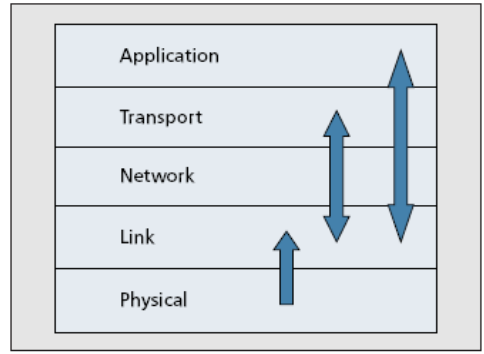

Figure 3. The cross-layer design approach [1]

As for joint source/channel coding, it has been shown that for video services over a CDMA channel with limited capacity, an effective way is to pass source significance information (SSI) from the source coder in the application layer to the channel coder in the physical layer. Thus, more powerful forward error correction (FEC) code (and therefore more overhead) can be used to protect more important information while no or weaker FEC may be applied to less important information. Such joint source/channel coding is a cross-layer approach, called unequal error protection (UEP). The authors have further argued that in case of capacity shortage, UEP schemes can result in more graceful quality degradation (and thus smaller distortion, or higher PSNR) than equal error protection (EEP). It has been shown that based on channel capacity, the optimal transmission rate and power allocation for packets of each priority can be found to minimize the average distortion of the received video by means of an optimization formulation over CDMA channels. This mechanism outperforms uniform power allocation, as it exploits the degree of freedom added by CDMA power allocation.

### 3.3 A Cross-Layer Scheduling Algorithm

Liu, Wang, and Giannakis have proposed a scheduling algorithm at the medium access control (MAC) layer for multiple connections with diverse QoS requirements, where each connection employs adaptive modulation and coding (AMC) scheme at the physical (PHY) layer over wireless fading channels [14]. A priority function (PRF) is defined for each connection admitted in the system. This priority function is updated dynamically depending on the wireless channel quality, QoS satisfaction, and services across all layers. The connection with highest priority is scheduled each time. Each connection is assigned a priority, which is updated dynamically based on its channel and service status. The connection with the highest priority is scheduled each time. At the MAC layer, each connection belongs to a single service class and is associated with a set of QoS parameters that quantify its characteristics. Following IEEE 802.16 standard, four QoS classes are provided: (i) unsolicited grant services (UGS), (ii) real-time polling service, (iii) non real-time polling service and (iv) best effort service. The unsolicited grant service supports constant bit rate (CBR) and fixed throughput connections and provides guarantees on latency, and jitter. Real-time polling service provides guarantees on throughput and latency but with more tolerance on latency as compared to unsolicited grant service. Non-real time polling service provides guarantees in terms of throughput only and is suitable for data applications, such as FTP. Best effort service provides no guarantee on delay or throughput and is used for HTTP and email applications. The designed scheduler has the following features:
- Efficient bandwidth utilization is achieved so that the scheduler does not assign a time slot to the connection with bad channel quality and multiuser diversity can be exploited.
- Delay bound is provided for applications that are based on real-time polling service.
- Throughput is guaranteed for non real-time polling service connections if sufficient bandwidth is available.
- Implementation complexity is low because the priority-based scheduler simply updates the priority of each connection per frame and allocates maximum time slots to the connection with the highest priority.

- The scheduler is flexible as it does not depend on any traffic or channel model.
- Scalability is achieved. When the available bandwidth decreases due to addition of new connections, the performance of connections with low-priority service classes will be degraded before high priority classes are admitted.

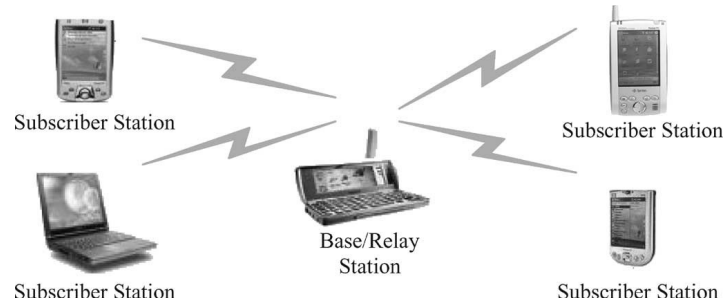

Figure 4: Network topology [14]

The architecture of the system is briefly described. Figure 4 shows the wireless networks topology, where multiple subscribers stations (SS) are connected to the base station (BS) or relay station over wireless channels. Multiple connections (sessions, flows) can be supported by each SS. All connections communicate with the BS using time division multiplexing (TDM) or time division multiple access (TDMA). The wireless link of each connection from the BS to each SS is depicted in Figure 5. A buffer is implemented at the BS for each connection and operates in first-in-first-out (FIFO) mode. The AMC controller follows the buffer at the BS (transmitter) and the AMC selector is implemented at the SS (receiver). At the PHY layer, multiple transmission modes are available to each user, with each mode representing a pair of specific modulation format and a forward error control (FEC) code. Based on the channel estimates obtained at the receiver, the AMC selector determines the modulation-coding pair (mode or burst profile), whose index is sent back to the transmitter through a feedback channel, for the AMC controller to update the transmission mode. Coherent demodulation and soft-decision Viterbi decoding are employed at the receiver. The decoded bit streams are mapped to packets, which are pushed upwards to the MAC.

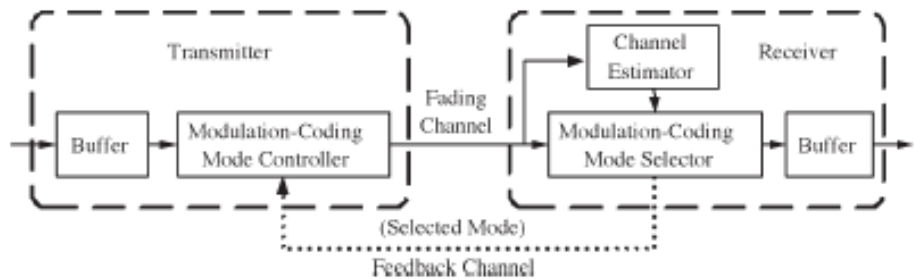

Figure 5: Wireless link from BS to SS [14]

### 3.4 A PHY-MAC Cross-Layer Optimization for Wireless Broadband Network

Triantafyllopoulou, Passas, and Kaloxylos have proposed a cross-layer optimization mechanism for multimedia traffic over IEEE 802.16 networks [15]. The scheme utilizes information provided by the PHY and MAC layers, such as signal quality, packet loss rate and the mean delay, in order to control parameters at the PHY and application layers and improve the performance of the system. Essentially, the adaptive modulation capability of the PHY layer and the multi-rate data-encoding feature of multimedia applications are combined to achieve an improved end-user QoS. The cross-layer optimizer is split into two parts- the BS part and the SS part, residing at the base station and the subscriber stations respectively. The BS part accepts an abstraction of layer-specific information regarding the channel conditions and QoS parameters of active connections, provided by the BS PHY and MAC layers. Based on this information, a specific decision algorithm determines the most suitable modulation and/or traffic rate of each SS, separately for each direction (uplink and downlink). Finally, the BS part informs the corresponding layers of the required modifications (Figure 6 (a)). If the decision of the BS part involves traffic rate changes, it communicates with the SS part through SS MAC layer, which instructs the SS application layer accordingly (Figure 6 (b)). The SS part may either accept the BS part's suggestions or refine them, based on its more accurate knowledge of the status of active connections. In the proposed architecture, the SS part is designed as a passive module that only instructs the SS application layer based on BS part's suggestion.

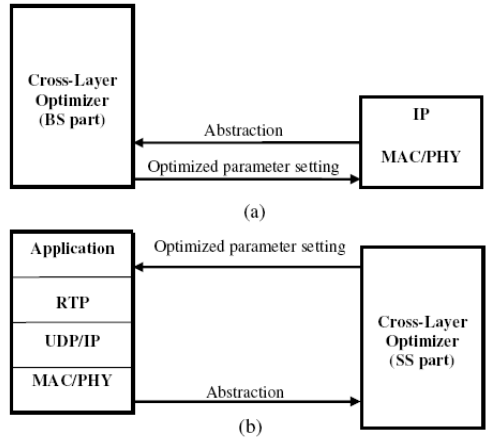

Figure 6: Functionality of cross-layer optimizer at BS (a) and SS (b) [15]

## 4 Adaptations at Different Layers of the Protocol Stack

Different types of adaptations are required at different layers of the standard protocol stack for providing a robust QoS support to multimedia applications over wireless networks. In Section 1, it has already been seen that wireless channels pose a number of challenges in this aspect. Considering the limitation of bandwidth in wireless systems, the most important target at the link layer is to increase link utilization. It is known that RTP/UDP/IP and TCP/IP have the problem of large header overhead on bandwidth-limited links. Header compression has been proven to be efficient for using those protocols. To handle the severe bandwidth and delay fluctuation in wireless Internet, available network condition estimation and congestion control are key issues needed to be addressed in the transport layer. Error protection, power saving, and proxy management are some of the important issues to be handled in the application layers. These layer-specific issues are described in details in the following subsections.

### 4.1 Link Layer Adaptation Mechanisms

There are several currently existing approaches for link layer adaptation under varying wireless channel conditions. The important ones in this category are: (i) application adaptive ARQ, (ii) priority-based scheduling, (iii) channel-aware scheduling. These are described in detail in the following.

*Application Adaptive ARQ:* To overcome packet loss, a technique called Automatic Repeat Request (ARQ) is used for packet retransmissions. ARQ uses acknowledgments and timeouts to achieve reliable data transmission. The receiver sends an acknowledgement (ACK) to the transmitter to indicate that it has correctly received a data frame or packet. The sender waits for a predefined period (timeout) for the ACK to arrive. If ACK arrives then the sender sends the next packet. Otherwise, it resends the earlier packet until it receives an ACK or exceeds a predefined number of retransmissions. ARQ can be implemented at the application/transport layer as well as the link-layer. Link-layer ARQ is more effective than application/transport layer ARQ because – (i) it has a shorter control loop and hence can recover lost data more quickly, {ii) it operates on frames that are much smaller than the IP datagrams and {iii) it might be able to use local knowledge that is not available to end hosts, to optimize delivery performance for the current link conditions. This information can include information about the state of the link and channel, e.g., knowledge of the current available transmission rate, the prevailing error environment, or available transmit power in wireless links [5].

However, optimal performance cannot be achieved using link-level ARQ as it may result in an undesirably large amount of data retransmission among different layers and consequent performance degradation in transport protocols. A more effective way of using the link-layer ARQ is to make it aware of the application QoS on a per packet basis [1]. The link-layer ARQ can then adjust its behavior accordingly. The effects of the adaptive ARQ are passed on to the application implicitly through packet drops and delay.

*Priority-based Scheduling:* In priority-based schedulers, packets are grouped into several classes with different priority according to their QoS requirements i.e. the MAC layer is made aware of the application layer QoS. Packets belonging to higher priority classes are more likely to be transmitted first. Packets in the same class are served in a FIFO manner. Based upon the priority scheduling mechanism, each QoS class will have some sort of statistical QoS guarantees [3].

*Channel-aware Scheduling:* In a multiple access wireless network, the radio channel is normally characterized by time-varying fading. To exploit the time-varying characteristic, a kind of channel-state dependent scheduling, called *multiuser diversity*, can be exploited to improve system performance (Section 3). For a wireless communication system with multiple MSs having independent time-varying fading channels, we can assume that the channels are either ON i.e. one packet can be transmitted successfully to the mobile user during the time-slot or OFF i.e. unsuitable for transmission. The scheduler at the BS MAC layer gets the channel state information from its PHY layer. The scheduler at the BS transmits to a user whose channel is in the ON state. In case more than one user channel is in ON state, the scheduler selects a user channel randomly. No data is sent by the BS when all the channels are OFF. For a 3-user case, all the channels will be in OFF state only for 1/8 of the time on average. Thus, total data rate achieved by the scheduler is (1-1/8) = 7/8 packets per slot. Hence average data rate per user is (7/8)/3 = 7/24 packets/slot. For round-robin scheduling with 3 users, each user will get 1/3 slot time. Since the user channels are equally likely to be ON or OFF in each timeslot, each user will get a data rate of (1/3)/2 = 1/6 packets/slot which is almost half that of the channel-aware multi-user diversity scheduler. Thus, overall resource utilization can be improved by using channel-aware scheduling mechanism [1, 6].

Due to different QoS metrics used in different layers, researchers propose to move the physical channel models up to link layer by converting physical layer QoS parameters into application-understandable QoS metrics. Wu and Negi have proposed *Effective Capacity* theory for modeling a wireless channel by two functions [26]. These two functions are: (i) the probability of nonempty buffer, and (ii) the QoS exponent of a connection that characterizes the queuing behavior in the link layer. In [2], EC model has been effectively used to estimate QoS parameters (e.g., delay bound, available bandwidth etc) of a multimedia application.

Zori, Rao, and Milstein have shown through analysis and simulation that a first-order Markov process is a good approximation model for data transmission over fading channels [27]. Following this approach, Zhang, Zhu, and Zhang have addressed the issues of resource allocation for scalable video transmission over 3G wireless networks [11]. The authors have first presented a method of estimation of time-varying wireless channels through measurements of throughput and error rate. Then a distortion-minimized bit allocation scheme with hybrid unequal error protection (UEP) and delay-constrained automatic repeat request (ARQ) that dynamically adapts to the estimated time-varying network conditions is proposed. The simulation results show that the proposed scheme can significantly improve the reconstructed video quality under degraded network conditions.

## 4.2 Transport Layer Adaptation Mechanisms

The wireless medium is very dynamic in nature due to the motion of the wireless devices, interference or fading. This fast changing, *small-scale channel variations* result in burst error at the receiver. In addition, there's a *large-scale channel variation* where the average channel state condition depends on users' locations and interference levels. Due to these channel variations, packet losses are inevitable at the receiver of the wireless communication system

In order to deliver multimedia over wireless networks, it's necessary to estimate the condition of the underlying network so that the strict QoS constraints for multimedia applications can be adhered to. Congestion may occur within a network when routers are overloaded with traffic which in turn causes their queues to build up and eventually overflow, leading to high delays and packet losses. Network conditions can be assessed mainly through congestion estimation based on - packet loss [1, 6] and current available bandwidth [3].

As mentioned in Section 3, TCP interprets all losses within a network as being congestion related. This is mainly because TCP was originally designed for wired networks with a reliable physical layer, where packet loss mainly results from network congestion. This characteristic of TCP is unsuitable for wireless networks since losses due to inherent channel errors are also treated as a signal of network congestion. This causes the source TCP to reduce its transmission rate by shrinking its congestion window size, even though there is no network congestion resulting in unnecessary decrease in throughput. In principle, packet loss due to channel errors should result in retransmissions not rate reduction. In order to improve the TCP performance in wireless scenario, it is necessary to differentiate the congestion-related packet losses from non-congestion packet losses. Two methods to achieve this objective - Snoop TCP and TCP with ECN, are described below.

*Snoop TCP:* Snoop TCP provides a reliable TCP-aware link layer. The mechanism is described in a scenario where data transfer occurs between a Fixed Host (FH) and a Mobile Host (MH) with a Base Station (BS) in between them. A snoop agent is created at the BS which buffers data at its link layer for retransmissions, instead of going back to TCP end points (FH and MH). Snoop maintains a state for each TCP connection traversing through the BS thus tracking TCP data and acknowledgements. Snoop caches unacknowledged TCP packets and uses the loss indications conveyed by duplicate acknowledgments and local timers to transparently retransmit lost data. It hides duplicate acknowledgments indicating wireless losses from the TCP sender, thereby preventing redundant TCP recovery. Snoop exploits the information present in TCP packets to avoid link layer control overhead. Snoop preserves end-to-end TCP semantics but it cannot work on encrypted datagrams. This makes it unsuitable in virtual private networks (VPNs).

*TCP with Explicit Congestion Notification (ECN):* ECN is an end-to-end mechanism to notify the sender whenever congestion occurs in the network. This method overcomes the inherent insensitivity of TCP congestion control mechanisms to delay or loss of individual packets focusing mainly on minimizing the impact of losses from a throughput perspective. Thus, TCP with ECN capabilities is tailor-made to improve the QoS of delay and packet–loss sensitive multimedia applications e.g., video-conferencing, VOIP etc. over wireless networks. In a standard IP packet header, an ECN field is included [8]. Whenever a router detects persistent (not transient) congestion in the network, it sets the ECN field and the packet is said to be marked. The marked packet eventually reaches the destination, which in turn informs the source about the congestion by setting the ECN-Echo flag in the TCP header. The source adapts its transmission rate accordingly using the usual TCP congestion control mechanisms of *slow-start*, *fast retransmit* and *fast recovery*. The ECN capability thus overrides any signal of packet losses as imminent congestion indication. This method requires the ECN scheme to be enabled at all (if any) intermediate router(s).

Raman, Balakrishnan, and Srinivasan have proposed an efficient transport layer protocol for transmission of images over wireless networks [23]. In the proposed protocol, named Image Transport Protocol (ITP), application data unit (ADU) boundaries are exposed to the transport module. This enables the transport module to perform out-of-order delivery of packets. As the transport is aware of application framing boundaries, the proposed approach expands on the application-level framing (ALF) philosophy, which proposes a one-to-one mapping from an ADU to a network packet or protocol data unit (PDU) [30]. ITP deviates from the TCP's notion of reliable delivery. Instead, it incorporates selective reliability, where the receiver is in control of deciding on what is to be transmitted from the sender at any instant. ITP runs over UDP, incorporates receiver-driven selective reliability, and uses a congestion manager (CM) to adapt to network congestion. It also enables a variety of new receiver post-processing algorithms such as error concealment that further improve the interactivity and responsiveness of reconstructed images.

Experimental results show that ITP is a very efficient and effective protocol for image transmission over loss prone wireless channels.

Yang, Zhang, Zhu, and Zhang have proposed an end-to-end TCP-friendly multimedia streaming protocol for wireless Internet which has been named WMSTFP (Wireless Multimedia Streaming TCP-Friendly Protocol) [19]. This protocol can effectively differentiate erroneous packet losses from congestive losses and filter out the abnormal round-trip time values caused by the highly varying wireless environment. Based on WMSTFP, the authors have proposed a novel loss pattern differentiated bit allocation scheme that applies unequal loss protection for scalable video streaming over wireless Internet. In particular, a rate-distortion-based bit allocation scheme has been proposed that considers both the wired and wireless network status to minimize the expected end-to-end distortion. The global optimal solution for the bit allocation scheme is obtained by a local search algorithm taking the characteristics of the progressive fine granularity scalable video into account. Analytical and simulation results have demonstrated the effectiveness of the mechanism. Figure 7 depicts the detailed diagram of the end-to-end scalable video streaming mechanism. The key components in this architecture consist of *WMSTFP congestion control*, *WMSTFP network monitor*, *Unequal loss protection (ULP) channel encoder*, and *loss differentiated rate distortion-based bit allocation*. WMFSTFP congestion control and WMSTFP network monitor provide network adaptation at end hosts, which mainly deal with probing and estimating the dynamic network conditions using the TCP-friendly protocol. The WMSTFP congestion control module adjusts sending rate on the sender side based on the feedback information, and the WMSTFP network monitor module on the receiver side analyzes the erroneous loss rate and congestive loss rate caused in a connection comprising both wired and wireless links and estimates the end-to-end available network bandwidth. The control data consisting of the estimated network bandwidth and other related network status parameters such as congestive packet loss rate, erroneous packet loss rate, and smoothed packet transmission time are fed back to the sender. Network-adaptive ULP channel encoder module protects different layers of progressive fine granularity scalable (PFGS) video against congestive packet losses and erroneous losses according to their importance and network status using Reed-Solomon (RS) codes [31]. Loss differential rate-distortion-based bit allocation module performs media adaptation control so that the total sending rate is adapted to the estimated network conditions. Based on the feedback information from the receiver, the bit allocation module on the sender side distributes the total sending rate between video bit rate and error protection rate according to the available bandwidth and different packet loss conditions in wired and wireless connections.

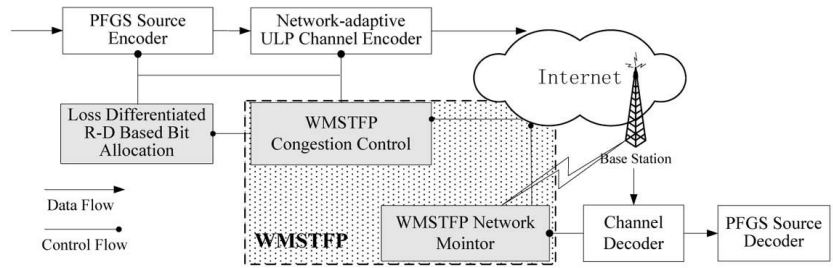

Figure 7: System architecture for scalable video streaming over wireless Internet [19]

Ahmed, Mehaoua, Boutaba, and Iraqi have proposed a media content analysis technique and a network control mechanism for adaptive video streaming over IP networks [24]. The proposed mechanism consists of three components: (i) a content-based video classification model for automatic translation from video application level QoS (e.g., MPEG-4 object descriptor and/or MPEG-7 meta-data framework) to network system level QoS (e.g. IP DiffServ per-hop-behaviors (PHBs)), (ii) a robust and adaptive application level framing protocol with video stream multiplexing and unequal forward error protection, (iii) a fine grained TCP-friendly video rate adaptation algorithm. The obtained experimental results shows that by breaking the OSI protocol layer isolation paradigm and injecting content-level semantic and service-level requirements within the transport and traffic control protocols, lead to intelligent and efficient support of multimedia services over complex network architectures.

Vieron and Guillemot have described a rate control algorithm that takes into account the behavior of TCP's congestion avoidance mechanism and the delay constraints of real-time streams [25]. Extending the TCP-Friendly Rate Control (TFRC) protocol [29], they have proposed a new protocol for estimating the bandwidth prediction model parameters. The protocol makes use of RTP and RTCP and takes into account the characteristics of multimedia flows (e.g., variable packet size, delay etc.). Based on the estimated current channel state, the states of the encoder and decoder buffers as well as the delay constraints of the real-time video source are translated into encoder rate constraints. This global rate control model coupled with a loss resilient video compression algorithm has been extensively experimented on various Internet links. Results have shown effectiveness of the protocol in terms of minimization of expected distortion and significant reduction in source timeouts.

### 4.3 Application Layer Adaptation Mechanisms

Due to real-time nature, multimedia services typically require QoS guarantees like large bandwidth, stringent delay bound and relatively error-free video/audio/speech quality. Multimedia services over the wireless channels become very challenging due to the dynamic uncertain nature of the channel resulting in variable available bandwidths and random packet losses. The main objectives of the application layer QoS control for multimedia communication over wireless networks are –(i) to avoid bursty losses and excessive delay (caused by network congestion) that have a devastating effect on multimedia presentation quality, and (ii) to maximize multimedia quality even when packet loss occurs in a wireless communication network. There have been a number of approaches currently existing in the literature in this regard. First, two important cross-layer approaches e.g., Joint Design of Source Rate Control and QoS-aware Congestion Control [7, 9] and Joint Design of Source Coding and Link Layer FEC/ Retransmission [1, 3] are described. Some other propositions are also described later.

*Joint Design of Source Rate Control and QoS-aware Congestion Control:* Congestion control for streaming media at the transport layer and source rate control at the application layer are employed to overcome the problems of multimedia communication over the wireless channels. In traditional layered design approach, source rate control and congestion control are designed independently and in isolation with each other. This imposes a limitation on the overall system performance e.g., end-to-end delay constraint and smooth playback quality. Congestion control for streaming multimedia usually needs to smooth its sending rate to help the application achieve smooth playback quality. However, this is not always possible as the source coding block at application layer can change the coding complexity and sending rate abruptly based on its QoS requirements, unless notified by the transport layer. Moreover, source rate control alone cannot guarantee the end-to-end delay constraint due to minimum bandwidth requirement and quality smoothness requirement, in the absence of congestion control mechanism at transport layer. Zhu, Zeng, and Li have proposed a joint source rate control and QoS-aware congestion control cross-layer mechanism to achieve a better overall system performance [7]. The architecture of the system is depicted in Figure 8. The mechanism works as follows. If the sending rate is made to temporarily violate the TCP-friendliness nature of the transport layer, the quality of the multimedia content is significantly improved. The long-term TCP–friendly sending rate is preserved by implementing the rate compensation algorithm [9]. At the application layer, the virtual network buffer management mechanism is used to translate the QoS requirements of the application in terms of the source and sending rates. There is a middleware component located between the application layer and the transport layer wherein the joint decision of the source rate and the sending rate is done. To make their proposed mechanism work effectively in wireless environment, the authors have incorporated the Analytical Rate Control (ARC) protocol presented in [10]. ARC protocol is intended to achieve high throughput and multimedia support for real-time traffic flows while preserving fairness to the TCP sources sharing the same wired link resources. The sender performs rate control using the ARC equation to avoid any unnecessary rate reduction due to wireless link errors, thus enabling the system to work properly in a wireless environment.

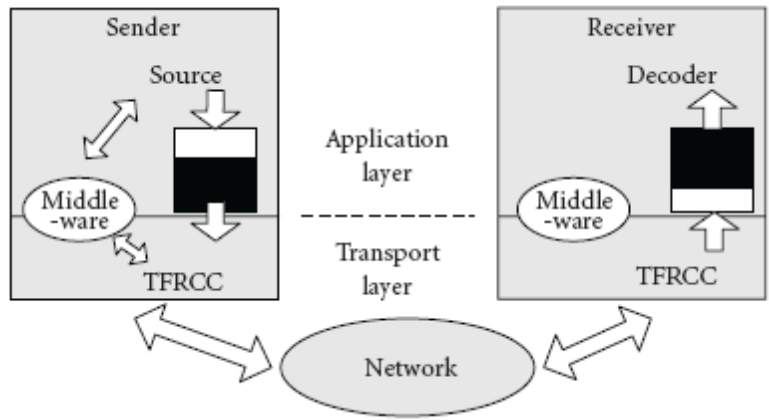

Figure 8: The system architecture for source rate control and congestion control [7]

*Joint Design of Source Coding and Link Layer FEC/Retransmission*: In order to adapt to the varying network conditions like loss, delay, variable bandwidth etc., the media codes are designed using scalable coding techniques. Scalability in video can be achieved by layered coding technique (e.g. in MPEG-4). The adaptation of audio codec, which also has a layered structure, can be achieved in a way similar to that of scalable video codec [16]. Speech codecs also allow dynamic rate adaptation, controlled by an in-band signaling procedure [3].

The layered coding technology divides the video into several layers and the incremental reception of layers increases the media fidelity. Video codecs encode a video sequence into one base layer and multiple enhancement layers based on any of the following three classes of layered coding techniques – temporal, spatial and signal-to-noise ratio scalability [17]. Different layers in a scalable coder have different importance in video transmission and reception. The correct decoding of the enhancement layers depend on the errorless receipt of the Base layer, Thus from the video reception point of view, Base layer is more important than Enhancement layers.

Forward error correction (FEC) and retransmissions are major link layer error correction mechanisms. FEC is a channel coding technique protecting the source data by adding redundant data during transmission. Thus FEC is not bandwidth efficient but very effective in applications with strict delay requirements such as voice communications (retransmission may induce huge latency). Applications where delay requirements are much relaxed, link layer retransmissions are more suitable as it is bandwidth efficient unlike FEC.

As mentioned earlier, in a multi-hop wireless scenario, packet losses can occur due to network congestion or wireless transmission errors, which invariably will have different loss patterns. According to [18], such different loss patters will get reflected as different perceived QoS at the application layer. A loss differentiated rate-distortion based bit allocation scheme is proposed in [19] that minimizes end-to-end video distortions taking the different loss patterns into account, which shows that both source coding and channel coding parameters can affect the final media quality. Joint Source and Channel Coding (JSCC) schemes are proposed to achieve the optimal end-to-end quality by adjusting the source and channel coding parameters simultaneously. As mentioned in Section 3, a simple JSCC scheme using UEP is presented by Jiang, Zhuang and Shen [1]. UEP can be performed with Bose-Chaudhuri-Hocquenghem (BCH) codes, Reed-Solomon (RS) codes, and Rate-Compatible Punctured Convolutional (RCPC) codes with different coding rates for packets with different priorities. A hybrid UEP scheme taking ARQ based retransmission based on the same SSI can also be implemented, where the base layer data can have maximum retransmissions with minimum or no retransmission at the enhancement layers. A delay-bound in such a hybrid scenario can be achieved by limiting the number of retransmissions [3].

In addition to the techniques described above, there are some other mechanisms currently proposed in the literature at the application layer. Superiori, Nemethova, and Rupp have proposed a robust error handling techniques for video streaming over mobile networks [20]. In order to avoid large overhead caused by smaller packets, the authors have proposed a scheme that utilizes the residual redundancy of the encoded video stream. At the decoder side there is a syntax analyzer that enables exact localization of errors within a packet. In addition, there is an entropy code resynchronization mechanism based on the out-of-band-signalized length indicators. It has been shown that the proposed approach provides substantial improvement in PSNR for the same rate, compared to the standard packet size reduction techniques.

Burza, Kang, and Van der Stock have described a robust streaming of combined MPEG audio/video content over in-home wireless networks, where the amount of data transmitted by the sender is dynamically adapted to the available bandwidth by selectively dropping data [21]. The bit-rate adaptation is achieved by using a packet scheduling technique called I-frame Delay (IFD) that performs priority-based frame dropping when the available bandwidth is limited. The proposed solution has been implemented using RTP (Real-time Transport protocol) and TCP (Transmission Control Protocol) at the transport layer.

Choi, Kellerer, and Steinbach have proposed a cross-layer optimization approach for wireless multi-user video streaming that jointly considers the application layer and the PHY/MAC layer of the protocol stack [22]. The optimizer maximizes the end-to-end QoS of the video streaming service jointly for all users while efficiently using the wireless resources. The authors have considered a video-streaming server located at the base station and multiple streaming clients. As in Figure 9, the clients are assumed to be sharing the same air interface and network resources but they request different video content. Necessary state information is first collected from the application layers and the radio link layer through the process of parameter abstraction. The process of parameter abstraction results in the transformation of layer specific parameters into parameters that are comprehensible for the cross-layer optimizer. The optimization is carried out with respect to a particular objective function. From a given set of possible cross-layer parameter tuples, the tuple optimizing the objective function is selected. After the decision on a particular cross-layer parameter tuple is made, the optimizer distributes the decision information back to the corresponding layers. The experimental results have demonstrated that for a small number of users and a small number of degrees of freedom in the optimization, significant quality improvements can be obtained.

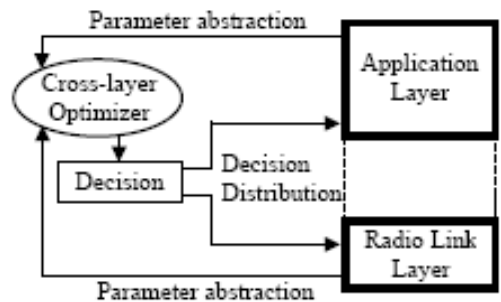

Figure 9: Cross-layer optimization architecture [22]

## 5 Conclusion

Cross-layer adaptations are essential for guaranteeing QoS supports in real-time multimedia traffic over wireless networks. In this paper, various adaptation mechanisms at the application layer, transport layer and link layer are discussed based on the currently existing propositions in the literature. More specifically, network-aware adaptive media source coding, dynamic estimation of the varying channel, adaptive and energy-efficient application and link-level error control, efficient congestion control, adaptive ARQ and priority-based scheduling are explicitly reviewed. However, cross-layer design is an extremely challenging task and lots of other issues need to be taken into consideration for an efficient design. QoS support in a multicast media streaming is one such area which requires attention [11]. Mobility of the users will bring in another dimension of complexity which will call for an efficient handling of the problem related to handoff while guaranteeing the application QoS. In mobile ad hoc networks (MANETs), changes in the topology of the network graph and the interference due to simultaneous communications will pose serious challenges too. Multi-path media streaming and QoS-aware MAC design are two cross-layer design approaches proposed in the literature for providing QoS support in MANETs [12,13]. However, any cross layer design should take a cautious and careful approach as some adverse impact on the system performance can occur in certain situations due to cross layer interactions [28]. Unbridled and extensive cross layer interactions can lead to a complex spaghetti design and thwart further innovations. Also such design will lack standardization and compatibility and portability features. This calls for a careful impact analysis and design of the cross layer protocol stack.